\documentstyle[aps,pre,multicol]{revtex}
\begin{document}
\title{Extracting and Stabilizing the Unstable State of Hysteresis Loop }
\author{Liu Yaowen,$^{1,}$\thanks{%
E-mail: chaosun@lzu.edu.cn} Ping Huican,$^1$ Zhao Hong,$^{1,2}$ Wang Yinghai$%
^{1,}$\thanks{%
Corresponding author. E-mail: wangyh@lzu.edu.cn. Fax: +86 931 8912172.}}
\address{$^1$Institute of Theoretical Physics, Department of Physics, Lanzhou\\
University, Lanzhou 730000, China \\
$^2$Department of Physics and Centre for Nonlinear Studies, Hongkong Baptist%
\\
Univerisity, Hong Kong, China}
\maketitle
\date{\today }

\begin{abstract}
A novel perturbation method for the stabilization of unstable intermediate
states of hysteresis loop (i.e. S-shaped curve) is proposed. This method
only needs output signals of the system to construct the perturbation form
without delay-coordinate embedding technique, it is more practical for
real-world systems. Stabilizing and tracking the unstable intermediate
branch are demonstrated through the examples of a bistable laser system and
delay feedback system. All the numerical results are obtained by simulating
each of the real experimential conditions.
\end{abstract}


\begin{multicols}{2}

\section{Introduction}

Bistability \cite{Gibbs1,Garmire,Lugiato,Hu1} has attracted much attention
not only in the experimental systems but also in the theoretical research in
the last twenty years for its possible application as an optical device and
switcher and so on. As a model problem in non-equilibrium statistical
mechanics \cite{Bonifacio}, bistability has also been considered as an
interesting subject. Usually, we get an S-shaped solution curve (i.e.
hysteresis loop) in bistable systems, where the upper and lower branches are
stable and the intermediate branch is unstable. When the mathematical model
of the system is known, the values of the intermediate may be calculated by
certain approaches. However, for the experimental systems in the absence of
a {\it priori} mathematical system model, getting the accurate positions of
the unstable intermediate state is experimentally inaccessible under
ordinary circumstances. How to get the unstable intermediate state is
considered as a particularly interesting subject recently \cite
{Zimm,Macke,Hu2}.

Zimmermann {\it et al}. \cite{Zimm} firstly obtained an unstable
intermediate branch of hysteresis loop (HL) in a thermochemical experiment
about the gaseous reaction mixture $S_2O_6F_2\rightleftharpoons 2SO_3F$,
even though they only achieved a part of the intermediate branch in the
experiment. Their method is to introduce an external delay feedback for a
system parameter to stabilize the unstable intermediate branch. As they
demonstrated, their method appeared suitable for only short delays; but for
longer delays in the external feedback, some periodic attractors (not
present in the original system) were predicted to exist by this method. So
in fact this external delay feedback control technique changes the original
system dramatically.

In 1990, Ott, Grebogi and Yorke (OGY) \cite{OGY} proposed a new method of
controlling a chaotic dynamical system by stabilizing one of the many
unstable orbit embedded in a chaotic attractor, only through small
time-dependent perturbations in some accessible system parameters. This
method was extended by Romeiras {\it et al}. \cite{Romeiras}. For an $n$%
-dimensional system 
\begin{equation}
{\bf z}_{i+1}=f({\bf z}_i,p),  \label{model}
\end{equation}
they used a linear feedback 
\begin{equation}
p_{i+1}=p_0+{\bf k}({\bf z}_{i+1}-{\bf z}_{*})  \label{law0}
\end{equation}
to stabilize the target orbit $z_{*},$ where ${\bf k,z}\in {\bf R}^n{\bf ,}$
and $p$ is one of the adjustable system parameters. OGY method has been
proved unable to change the original system. Later researchers extended this
method to controlling the unstable state by using time delay coordinates
technique \cite{Dressler}. Principally, OGY method can also stabilize the
unstable intermediate branch of hysteresis loop (HL), but a potential
difficulty with a nonadaptive application of this technique to real-world
system is that the unstable state must be known before the use of this
method. For experimental, especially ``black-boxes'' systems (i.e. little or
no information exists regarding their governing equation), it is very
difficult to determine the unstable state $z_{*}$ exactly. When the OGY
feedback (\ref{law0}) invalidates in the experiment systems, Z. Gills {\it %
et al. }\cite{Gills}{\it \ }used $p(t)=p_0+k[z(t)-z_{ref}],$ ( $z_{ref}$ is
a ``guess'' value for the control objective) and S. Bielawski {\it et al. }%
used $p(t)=p_0+\beta \dot{z}$ in Ref. \cite{Bielawski} respectively, to
stabilize the unstable states. Some of us addressed another method based on
the OGY method in the variable-parameter space to control chaos in Ref. \cite
{Zhao1}. We suggested a feedback form 
\begin{equation}
p_{i+1}=p_0+{\bf k}({\bf z}_{i+1}-{\bf z}_i)+k^{\prime }(p_i-p_0),
\label{law1}
\end{equation}
where ${\bf k}$ and $k^{\prime }$ are control coefficients. It is different
from the previous tracking techniques \cite{Gills,Shinbrot} since it need
not ``guess'' or ``predict'' the control target $z_{*}$ and the unstable
state in the system can be stabilized automatically by choosing suitable
control coefficients.

In 1993, Macke {\it et al}. \cite{Macke} used a continuous feedback method
to acquire the unstable intermediate state of HL{\it \ }in an optical
bistable system. As a challenge for the high-dimensional systems
(spatiotemporal systems), Hu and He \cite{Hu2} numerically stabilized
unstable state of bistability in a spatiotemporal system described by a
partial differential equation with infecting negative feedback. Recently, we
successfully expanded our previous method \cite{Zhao1} to stabilize the
unstable state existing in a bistable system \cite{Zhao2} and simultaneously
pointed out that there exists a transformational law about control
coefficients between the stable and unstable states by using feedback form (%
\ref{law1}). Based on the Ref. \cite{Zhao2}, here we show a more applicable
method to control the unstable intermediate state in bistable systems.

This paper is organized as follows: In Sec. II, the basic idea of the
control method is shown and some advantages of the technique are also
presented. In Sec. III, we demonstrate the effectiveness of our technique by
applying it to five- and infinite-dimensional bistable systems respectively.
At the end of the paper, a summary of this paper and the prospect of
applications for experimental systems are discussed.

\section{The basic idea of the control method and its advantages}

Most experimental systems are essentially ``black-boxes'' for which one can
only get a series of output signals ${\bf z}_1{\bf ,z}_2{\bf ,z}_3{\bf ,...,}
$ and measure some adjustable system parameters, here ${\bf z}_i\in {\bf R}%
^n $. Some researchers have stabilized unstable states in some experimental
systems by using a delay-coordinate embedding technique \cite{Dressler}.
Here we illustrate a well-suited method for experimental systems to control
the unstable steady states without this embedding technique.

To simplify the analysis we first consider a one-dimensional time series,
which can be described by a discrete dynamical system as follows 
\begin{equation}
x_{i+1}=f(x_i,p,c),  \label{1}
\end{equation}
where $x\in R$, and $f$ is unknown but is sufficiently smooth. Here $p$ is
considered as the parameter that is available for external adjustment but is
restricted to lie in some small interval, $\left\| p-p_0\right\| <\delta ,$
around a nominal value $p_0$. $c$ is the bifurcation parameter (maybe $p$
and $c$ are the same parameter). Let $x_{*}(p_0)$ be a fixed point of (\ref
{1}) (i.e., controlling objective). For the one-dimensional system, the
feedback form (\ref{law1}) can be chosen as 
\begin{equation}
p_{i+1}=p_0+k(x_{i+1}-x_i)+k^{\prime }(p_i-p_0)  \label{2}
\end{equation}
to stabilize and track $x_{*}$, where $k$ and $k^{\prime }$ are control
coefficients.

Hysteresis loop is a typical bistable phenomenon as sketched in Fig 1(a). In
this figure, $x_s$ denotes the stable lower and upper branches and $x_u$ is
the unstable intermediate branch. The two turning points of the stable and
unstable states are tangential bifurcation points. $c_b$ denotes the value
of bifurcation parameter at the tangential bifurcation point and $x_b$
represents the corresponding value of a variable. $c_1$ is a neighboring
parameter value of tangential bifurcation point. $x_{c_1}$ denotes the value
of variable located in the stable lower branch at $c=c_1$ and $%
x_{c_1}^{\prime }$ is the corresponding value located unstable intermediate
branch. For a hysteresis loop, we have developed a method in Ref. \cite
{Zhao2} to stabilize the unstable intermediate state by using the
characteristic of the tangent bifurcation. When applying the feedback law (%
\ref{2}) to a bistable system, there exists a transformational law between
the stability region of $x_{c_1}$ and the stability region of $%
x_{c_1}^{\prime }$ in the control coefficients space (for detail see Ref. 
\cite{Zhao2}). We have proved that the two regions have no intersection, as
sketched in Fig. 1(c). One can target the unstable intermediate branch of HL
according to the method in Ref. \cite{Zhao2} from which we derived a
transformational law turning the coefficients obtained at the stable state
into those suitable for stabilizing the unstable state, but it is complex
because a {\em matrix} must be calculated to get the control coefficients $k$
and $k^{\prime }$ of stable state. To avoid the difficulty, here we present
a very easy method to get the suitable $k$ and $k^{\prime }$ without
calculation.

Firstly, to determine $k^{\prime },$ we have proved the fact in Ref. \cite
{Zhao2} that the two stability regions of $x_{c_1}$ and $x_{c_1}^{\prime }$
are separated by a supercurve $k^{\prime }=1$ and the stability region of $%
x_{c_1}$ is located below the supercurve while that of $x_{c_1}^{\prime }$
is located above when $c=c_1$ [see Fig. 1(c)]. $k^{\prime }=1-\delta $ is
located in the stability region of $x_{c_1},$ and $k^{\prime }=1+\delta $
lies in the region of $x_{c_1}^{\prime }$, where $\delta $ is a small
positive number. If one applies $k^{\prime }=1-\delta $ to the system, $%
x_{c_1}$ is stable while $x_{c_1}^{\prime }$ is unstable. On the contrary
when one uses $k^{\prime }=1+\delta $ to the system, $x_{c_1}$ becomes
unstable and $x_{c_1}^{\prime }$ becomes stable. So we can easily choose $%
k^{\prime }=1\pm \delta $ to stabilize the unstable or stable state.

Secondly, in order to determine $k$, one should accept the fact that the
point $(k,k^{\prime })=(0,0)$ in the control coefficient space must be
located within the stability region of $x_{c_1},$ as sketched in Figs. 1(c)
and (d). It is easy to understand that because in this case $p_{i+1}$ is $%
p_0 $ according to the feedback form (\ref{2}), the system must stabilize
the $x_{c_1}$ state naturally. At the tangential bifurcation point $c_b,$
the origin of coordinate is located at the boundary of the region [Fig.
1(b)]. Thus one can reasonably expect that the stability region of $x_{c_1}$
must lie in the neighborhood of the point $(k,k^{\prime })=(0,0)$ and can be
easily found by scanning the coefficient space around the origin point $%
(0,0),$ [see the down triangle in Fig. 1(c)]. One can take the values of $k$
and $k^{\prime }$ in the region of $x_{c_1},$ say $A(k_1,1-\delta )$, where $%
k_1$ should be chosen in the middle value of region in the direction of $k$
at $k^{\prime }=1-\delta $ in order to assure that $A^{\prime }(k_1,1+\delta
)$ is also located in the stability region of $x_{c_1}^{\prime }$. If $%
(k,k^{\prime })=(k_1,1+\delta )$ is used to replace $(k_1,1-\delta ),$ the
perturbations (\ref{2}) can drive the system dynamical trajectories to
depart from the stable lower state $x_{c_1}$ and terminate on the unstable
intermediate one $x_{c_1}^{\prime }$ at $c_1$ automatically. In this case
the lower branch loses stability and the unstable intermediate state becomes
stable. Taking $(k,k^{\prime })=(k_1,1+\delta )$, one can track the
intermediate branch when $c$ decreases.

Finally, if the control coefficients $(k,k^{\prime })=(k_1,1+\delta )$ lose
control ability when bifurcation parameter $c$ decreases at $c=c_2$ [see
Fig. 1(d)], one can use $k_2$ to replace the control coefficient $k_1,$ and
apply it to the system according to the feedback law (\ref{2}) to stabilize
the unstable intermediate state continuously. The reason is demonstrated in
the following. As sketched in Figs. 1(c) and (d), with the bifurcation
parameter $c$ being far away from the tangential bifurcation point $c_b,$
the stability regions of $x_s$ and $x_u$ are also moving along in the
positive or negative direction of $k$ (for different systems, the moving
direction is different. Without loss of generality, we suppose the region
moves to the negative direction of $k$ in Fig. 1). At the same time, the
size of regions is also changed, but the origin of coordinates $(0,0)$ is
always located in the stability region of $x_s$ (the reason has been
demonstrated in the last paragraph). At $c=c_2,$ though $A(k_1,1-\delta )$
is still located in the stability region of $x_{c_1}$ (the down triangle), $%
A^{\prime }(k_1,1+\delta )$ does not lie in the stability region of $x_{c_2}$
(the up triangle), so $A(k_1,1+\delta )$ is out of control for the
intermediate state. However, $B^{\prime }(k_2,1+\delta )$ from the
corresponding point $B(k_2,1-\delta )$ is located in the stability region of 
$x_{c_2}$, thus it can continuously stabilize and track the intermediate
branch. This step can be repeated with the further decrease of parameter $c$
until the whole intermediate branch of the HL is obtained{\it . }Generally
(in the following applications of this paper), only by choosing $k_1$ and $%
k^{\prime }$ at $c_1,$ one can get the whole intermediate state of
hysteresis loop by this technique. This just verifies the robustness of our
method (for detail see Sec. IIIA).

For an $n$-dimensional system ${\bf z}_{i+1}=f({\bf z}_i,p,c)$ with several
unstable directions, where ${\bf z}\in {\bf R}^n,$ one can use 
\begin{equation}
p_{i+1}=p_0+{\bf k}({\bf z}_{i+1}-{\bf z}_i)+k^{\prime }(p_i-p_0)
\label{aaa}
\end{equation}
to control the unstable intermediate state of HL, here ${\bf k\in R}^n.$ We
can choose $k_i(k_i\in {\bf k,}$ $i=1,2,3,...n)$ according to the method of
choosing $k$ in a one-dimensional case. $k^{\prime }$ can still be chosen as 
$1-\delta $ for the stable state and $1+\delta $ for the unstable one. In
the examples of Sec. III, we consider two dynamical systems which
instability is characterized by only one unstable direction. Thus the
stabilization of the unstable intermediate state of HL only requires one
system variable to be monitored and one system parameter perturbed, so only
two control coefficients $k$ and $k^{\prime }$ are used, as done in a one
dimensional case, to realize the control.

Usually, for the systems whose mathematical model are known, the unstable
intermediate steady state of HL can also be achieved numerically by Newton's
method, with the help of the priori knowledge about state functions suitable
to the dynamical system. But for the experimental systems whose governing
equations unknown, there is usually no way to know the state functions, so
the Newton's method will invalidate. Nevertheless, the approach in the
present paper whose goal is to provide a technique for the experimental
systems to stabilize the unstable intermediate state of hysteresis loop,
stabilizes the unstable state by making small perturbations to an accessible
system parameter, thus it is well-suited for the real-world experimental
system.

In addition, this method has other advantages: (i) It can leave the original
system unchanged because it is an extended version of OGY method and
inherits its characteristic. (ii) The feedback form (\ref{2}) develops OGY
method since one does not need to know (``guess'' or ``predict'') the
desired orbit $x_{*}$ (which, in fact, is difficult to obtain in
experiments), and what one needs only to do is to record output signals of
two times in succession (i.e. $x_i,x_{i+1}$) and the parameter $p_i$ in this
step. Thus one can obtain the parameter value $p_{i+1}$ for the next time to
control the system according to the feedback form (\ref{2}). It is not
difficult to design the corresponding experiment elements based on the
original system, where $k$ and $k^{\prime }$ can be considered as adjustable
signal amplifiers. When the control coefficients $k$ and $k^{\prime }$ are
chosen suitably, the system will be stabilized onto the control target by
itself. (iii) Using the outputs of the system to construct the control law
without the delay-coordinate embedding technique, one does not need to know
the characteristic of tangent space and does not need to calculate the
derivative matrix of Eq. (\ref{1}) to get the control coefficients. This
gives us a convenient way in the experimental system in which mathematical
model is absent. (iv) This technique has good resistance to noise. Some
experimental systems are influenced by noise (e.g. some system parameters
change as a function of time influenced by temperature, etc.), thus the
position of the desired orbit $x_{*}$ moves a small distance due to the
change of system parameters. Since Eq. (\ref{2}) does not include $x_{*}$
explicitly, the perturbations with any point in the stability region can
force the system trajectories to evolve towards the exact position of $x_{*}$
(interested readers are referred to Refs. \cite{Zhao1,Zhao2}).

\section{Application in bistable systems}

\subsection{Five-dimensional model: An optical bistable system.}

Now, let us consider an optical cavity filled with a passive medium,
containing of homogeneously broadened two-level atoms, and driven by an
external coherent laser field \cite{Lugiato,Hu1,Segard}. Considering only
the single mode case, applying the plane wave approximation and taking the
mean-field limit, one can reduce the Maxwell-Bloch equation to 
\begin{eqnarray}
\dot{x} &=&-\tilde{\kappa}[(1+i\theta )x-y+2Cp,  \nonumber \\
\dot{p} &=&xD-(1+i\tilde{\Delta})p,  \label{model2} \\
\dot{D} &=&-\tilde{\gamma}[(x^{*}p+xp^{*})/2+D-1],  \nonumber
\end{eqnarray}
where the variables $x,p$ are the complex output field and the atomic
polarization. $D$ is the real population difference. Thus the equations are
essentially five dimensional. The system parameters $C,\tilde{\kappa},$ and $%
\tilde{\gamma}$ respectively represent the small-signal atomic gain, the
cavity linewidth, and the atomic population decay rate. Both $\tilde{\kappa}$
and $\tilde{\gamma}$ are scaled to the homogeneous linewidth $\gamma _{\bot
} $ [i.e. $\tilde{\kappa}=(\kappa /\gamma _{\bot }),\tilde{\gamma}=(\gamma
_{\Vert }/\gamma _{\bot })$]. Given the frequencies of the external field,
the atoms, and the cavity as $\varpi _0,\varpi _a,$ and $\varpi _c$, we
scale the atomic detuning and the cavity mistuning as $\tilde{\Delta}%
=(\varpi _a-\varpi _0)/\gamma _{\bot }$ and $\theta =(\varpi _c-\varpi
_0)/\kappa .$ The time $t$ is measured in units of polarization relaxation
time, $\gamma _{\bot }^{-1}$. $y$ is the amplitude of the external field, an
adjustable system parameter.

Under appropriate conditions, the system exhibits a bistable steady state
behavior, and the transmitted light varying discontinuously with the
incident field shows a hysteresis loop \cite{Lugiato}. Using the
fourth-order Runge-Kutta integration and with fixed step size $dt=0.01,$ we
get a series of output signals about the output field $x$ from Model (\ref
{model2}). We numerically plotted the HL of this system governed by Eq. (\ref
{model2}) at $C=30,\theta =20,\tilde{\gamma}=2,\tilde{\Delta}=4,$ and $%
\tilde{\kappa}=0.5,$ as illustrated in Fig. 2(a). These parameters can be
realized in the experiment of Ref. \cite{Segard}. Clearly, when the
bifurcation parameter $y$ increases from zero to $11$, the real part of
output variable $x$ (i.e. Re$x$) jumps up to the upper steady state at $%
y=9.43.$ Once on the upper branch, if the bifurcation parameter is slowly
decreased to zero, Re$x$ falls down to the lower steady state at $y=7.0.$
Conventional stability analysis finds the upper and the lower branches of
the S-shaped curve to be stable. The intermediate branch is found to be
unstable and is experimentally inaccessible under ordinary circumstances.
When the adjustable parameter $y=9.42,$ which lies in the neighborhood of
the turning point from the lower to the upper state, we can easily get the
stability region $k\in (-28.4,-8.9)$ at $k^{\prime }=0.98$ by using control
law $y_{i+1}=y_0+k($Re$x_{i+1}-$Re$x_i)+k^{\prime }(y_i-y_0)$ in the system.
In our work, we use $k=-16$ which lies in the middle of the stability
interval, apply $k^{\prime }=1.02$ to replace $k^{\prime }=0.98$ to
stabilize the unstable intermediate state, and we find the system stabilizes
automatically to the (original unstable) intermediate state of HL. Figure
2(b) shows the targeting process versus integral times $n$. At the beginning
of the process, Re$x$ is located in the stable lower state without control;
at $n=12000,$ we begin to add a control with $k=-16,k^{\prime }=1.02$ to the
system, we find Re$x$ increases slowly and is completely located in the
intermediate (original unstable) state at $n=19000;$ when $n=40000,$ we use $%
k^{\prime }=0.98$ to replace $1.02$ and $k$ is still $-16$, we observe that
Re$x$ decreases and turns back to the stable lower state at $n=60000.$ In
fact, when the system lies in the intermediate state, we also cancel the
control and wait the system to restore to the lower state slowly. When
system stabilizes the intermediate state at $y=9.42,$ then decreasing the
bifurcation parameter $y$ from $9.42$ to $7.0$ and using the control law (%
\ref{2}) with $k=-16,k^{\prime }=1.02$ continuously, one can stabilize the
whole unstable intermediate state of HL, as shown in Fig. 2(a). Without
modifying the control coefficients, we can stabilize and track the whole
intermediate state. Obviously, this technique is robust for the experimental
systems.

\subsection{Infinite dimensional model: A delay feedback differential system.
}

The method above can also be used in the experimental system designed by
Gibbs and Hopf {\it et al}. (see Fig. 1 of Ref. \cite{Gibbs}). The
delay-line hybrid optical bistable device can be described by a delay
differential equation with an infinite number of degrees of freedom, 
\begin{equation}
\tau ^{\prime }\dot{\varphi}(t)=-\varphi (t)+A^2\{1+2B\cos [\varphi
(t-t_R)-\varphi _0]\}  \label{model3}
\end{equation}
where $\varphi =\pi V/V_H,V$ is the voltage fed to the modulator, $V_H$ is
the half-wave voltage of the modulator, $\tau ^{\prime }$ is the composite
detector-feedback-PLZT response time, $B$ is a coefficient that measures the
ability of the modulator to achieve extinction between the crossed
polarizer, $\varphi _0=-\pi /2$ can be set in the experiment, the
bifurcation parameter $A^2=\pi \mu =$ $CG_1G_2I_{in}$ is proportional to the
gains $G_1$ and $G_2$ of the amplifiers and to the input laser intensity $%
I_{in}$, which is the adjustable parameter in their experiment. $\mu $ can
be measured directly by breaking the beeking the feedback loop between the
final amplifier and the modulator. Let $t=t_R\bar{t},\bar{\varphi}(\overline{%
t})=\varphi (t)$ to rescale the Eq. (\ref{model3}), we then have 
\[
\frac{\tau ^{\prime }}{t_R}\stackrel{\cdot }{\bar{\varphi}}(\bar{t})=-\bar{%
\varphi}(\bar{t})+\pi \mu \{1+2B\cos [\bar{\varphi}(\bar{t}-1)-\varphi
_0]\}. 
\]
Denoting $\tau =\tau ^{\prime }/t_R$ and dropping the bars from $\overline{t}%
,\overline{\varphi }$, we get 
\begin{equation}
\tau \dot{\varphi}(t)=-\varphi (t)+\pi \mu \{1+2B\cos [\varphi (t-1)-\varphi
_0]\},  \label{model3a}
\end{equation}
where $B=0.3,$ as be chosen in Ref. \cite{Gibbs}.

Equation (\ref{model3a}) can be solved by using Adams interpolation
numerically. In order to trace the course of evolution, one can draw the
evolution curve of the variable $\varphi (t)$ via the time $t.$ We plot the
hysteresis loop at $\tau =6.25,$ $\mu $ from 0 to 3.0 and then back to 0, as
shown in the Fig. 3(a). This is under the corresponding experimental
condition of Gibbs {\it et al.} [i.e. $t_R=160\mu s\ll \tau ^{\prime }=1ms$
in Fig. 1(b) of Ref. \cite{Gibbs}]. In Fig. 3(a), $\varphi _i$ is obtained
by measuring $\varphi (t)$ in every time unity. In the increase of $\mu $,
the lower branch loses stability at $\mu =1.42$ and the system jumps up to
the upper branch. In the opposite direction, the upper branch loses
stability at $\mu =0.9$ and the system falls down to the lower branch. There
exists a bistable region from $\mu =0.9$ to $\mu =1.42.$ Obviously, our
numerical result is in excellent qualitative agreement with their
experiment. However, the intermediate branch of hysteresis loop [dotted line
in the Fig. 3(a)] has not been observed in their experiment. If the control
law $\mu _{i+1}=\mu _0+k(\varphi _{i+1}-\varphi _i)+k^{\prime }(\mu _i-\mu
_0)$ is added to the system (we add a control every time delay unit) at $\mu
_0=1.40$, here $k=-0.6,k^{\prime }=1.02,$ the lower stable branch will be
found to lose stability and the system will be locked to the intermediate
unstable branch ultimately. At this time, if the control is cancelled or $%
k=-0.6,k^{\prime }=0.98$ used to renew the control coefficients, the
intermediate state is observed to lose stability and the system turns in the
end back to the lower branch via the time $t,$ as shown in Fig. 3(b). When
the system stabilizes the intermediate branch by $k=-0.6,k^{\prime }=1.02$
at $\mu _0=1.40$, if one decreases the bifurcation parameter $\mu $ from $%
\mu _0=1.40$ to $\mu _0=1.0$, one will get the whole unstable intermediate
state of the HL, as presented in Fig. 3(a).

We also achieve the whole unstable intermediate state successfully in
another parameter, shown in Fig. 4. From this figure, one can find that the
upper branch of the bistable region has been in the self-pulsing region, but
we can also scan the intermediate unstable branch by the control method in
this paper.

\section{Conclusion and discussion}

In summary, we have numerically achieved the unstable intermediate states of
hysteresis loop in two bistable systems by a practical control method. Using
the feedback law $p_{i+1}=p_0+k(x_{i+1}-x_i)+k^{\prime }(p_i-p_0)$ in a
bistable system with hysteresis loop, the stability regions of stable state $%
x_s$ and unstable intermediate state $x_u$ have no intersection in $%
k-k^{\prime }$ plane \cite{Zhao2}. The two regions are separated by a
supercurve $k^{\prime }=1$ and the stability region of $x_s$ is located
below while that of $x_u$ is located above. For different systems, the
stability interval of $k$ is different but which can be easily obtained by
scanning the neighbor of zero while the value of $k^{\prime }$ is same and
can be chosen as $1+\delta $ for the unstable states. One of the main
advantages of this targeting method is that as the targeting procedure
proceeds the stable state becomes unstable while the unstable one (i.e., the
targeting objective) becomes stable and therefore one need not ``guess'' (or
``predict'') the position of the unstable state in advance, which certainly
benefits the experimentalists.

On the other hand, this method need not reconstruct the system by
delay-coordinate embedding technique or know the characteristic of the
tangential space of the system, and what one only need do is to measure the
output signals. From output signals, one can construct the control law
easily. Because of this simplicity, the approach is of practical interest
for the stabilization of intermediate branch in bistable experiment systems.
For the experiments, the key point is to design a group of experiment
elements to complete perturbation for the adjustable parameter $%
p_{i+1}=p_0+k(x_{i+1}-x_i)+k^{\prime }(p_i-p_0)$ for the system. We should
like to emphasize that this method may be applied by experimentalists to
realize the unstable output related to bistability from the experimental
device even if the global mathematical model of the system is unknown. The
authors of this paper expect that it will be verified in optical or other
bistable systems such as electronic circle, chemical system, magnetic
research field, and others.

\begin{center}
ACKNOWLEDGMENT
\end{center}

This work was supported in part by the Natural Science Foundation of China,
and in part by Doctoral Education Foundation of National Education a
Committee and the Natural Science Foundation of Gansu Province.

\end{multicols}
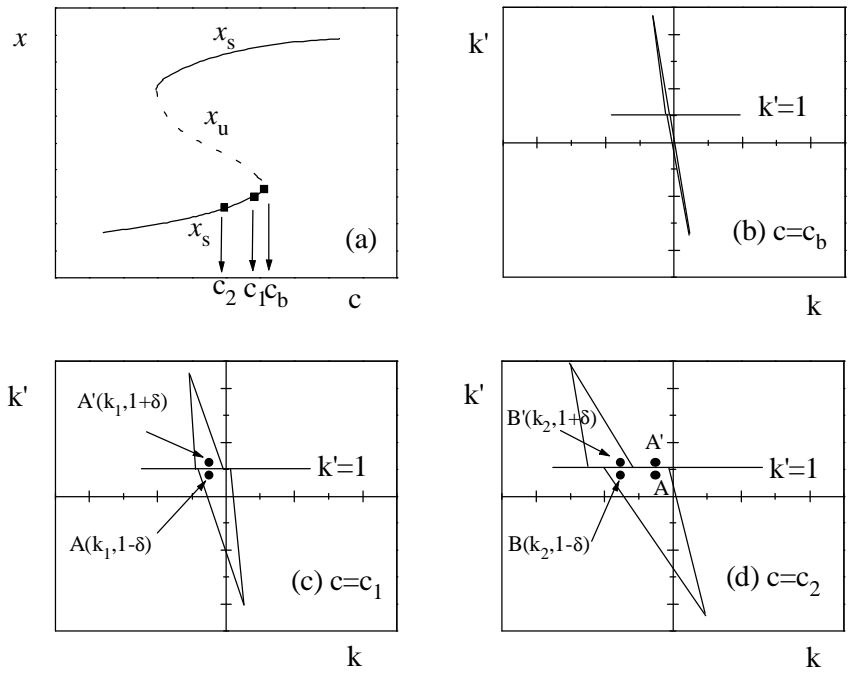
\begin{figure}[tbp]
\caption{(a). Schematic of hysteresis loop. $x_s$ denotes the stable branch
and $x_u$ denotes the unstable intermediate branch. $c$ is a bifurcation
parameter and $c_b$ is the tangential bifurcation point. (b)-(d) are the
schematices of the stability regions of $x_s$ (the solid-line down triangle)
and the stability regions of $x_u$ (the solid-line up triangle) at different
values of $c$ in the k-k' plane.}
\label{fig1}
\end{figure}

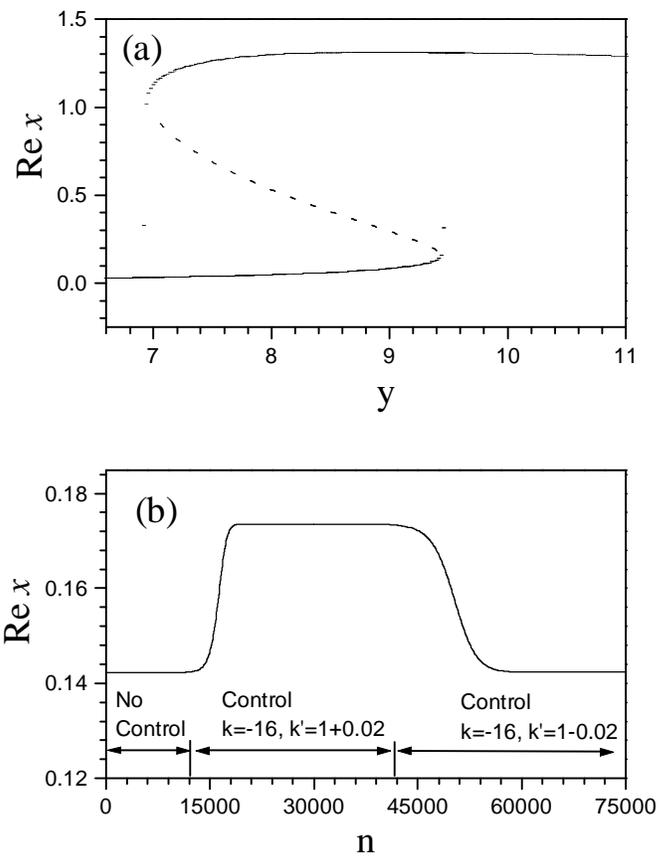
\begin{figure}[tbp]
\caption{Controlling of the unstable intermediate branch. (a). Hysteresis
loop of an optical bistable system and the controlling result of the
inermidiate branch. Bar scatters show the lower and upper stable states, and
the dotted line denotes the intermediate unstable state. (b) Re$x$ versus
integral times $n$ for a representative control technique trial. System
switches between the lower stable and the intermediate states by using $%
k=-16, k^{\prime}=1\pm 0.02.$}
\label{fig2}
\end{figure}

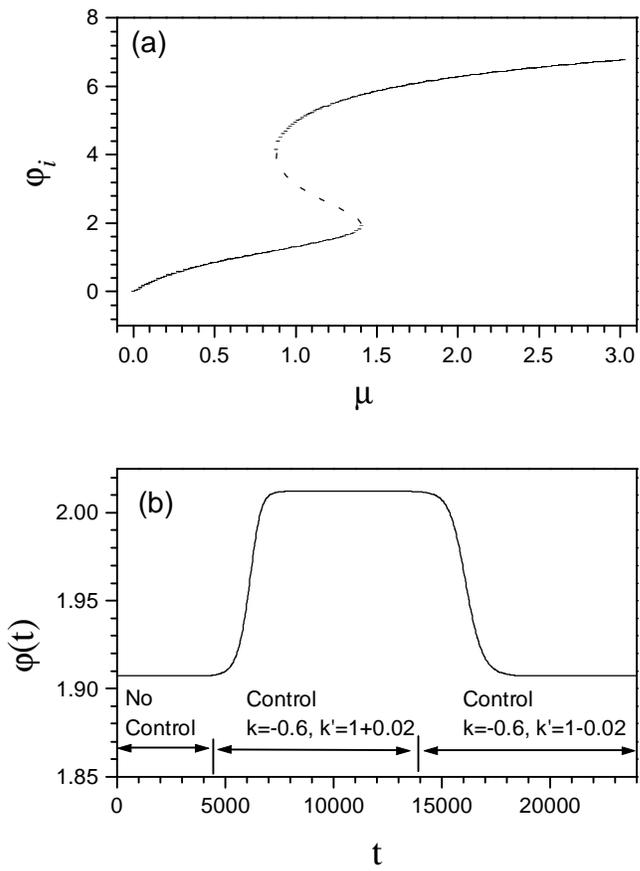
\begin{figure}[tbp]
\caption{(a) Hysteresis loop at $\tau=6.25$. Bar scatters show the lower and
upper stable states, and the dotted line denotes the intermediate unstable
state. (b) $\varphi (t)$ versus time for a representative control techenique
trial. The system switches diagram between the lower stable and intermediate
unstable states by using $k=-0.6, k^{\prime}=1\pm 0.02.$ }
\label{fig3}
\end{figure}

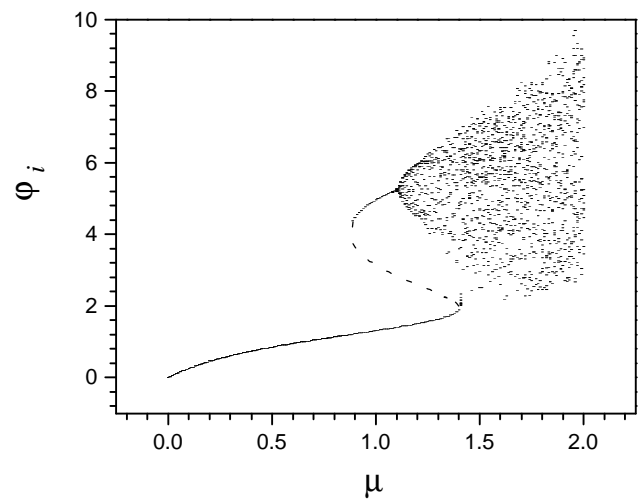
\begin{figure}[tbp]
\caption{Hysteresis loop at $\tau =0.2$. The bar scatters and dotted line as
same as Fig. 3(a). }
\label{fig4}
\end{figure}


\begin{references}
\bibitem{Gibbs1}  H. M. Gibbs, S. L. McCall, and T. N. C. Venkatesan, Phys.
Rev. Lett. {\bf 36} (1976) 1135.

\bibitem{Garmire}  E. Garmire, J. H. Marburger, and S. D. Allen, Aply. Phys.
Lett. {\bf 32} (1978) 320.

\bibitem{Lugiato}  L.A. Lugiato, L.M. Narducci, D.K. Bandy, C.A. Pennise,
Opt. Comm. {\bf 43} (1982) 281.

\bibitem{Hu1}  G. Hu, G. J. Yang, Phys. Rev. {\bf A38} (1988) 1979.

\bibitem{Bonifacio}  R. Bonifacio and L. A. Lugiato, Phys. Rev. Lett. {\bf 40%
} (1978) 1023.

\bibitem{Zimm}  E. C. Zimmermann, M. Schell, and J. Ross, J. Chem. Phys. 
{\bf 81} (1984) 1327.

\bibitem{Macke}  B. Macke, J. Zemmouri, and N. E. Fettouhi, Phys. Rev. {\bf %
A47} (1993) R1609.

\bibitem{Hu2}  G. Hu, K.F. He, Phys. Rev. Lett. {\bf 71}(1993) 3794.

\bibitem{OGY}  E. Ott, C. Grebogi, and J. A. Yorke, Phys. Rev. Lett. {\bf 64}
(1990)1196.

\bibitem{Romeiras}  F. J. Romeiras, C. Grebogi, E. Ott, and W. P. Dayawansa,
Physica {\bf D58 }(1992) 165.

\bibitem{Dressler}  U. Dressler and G. Nitsche, Phys. Rev. Lett. {\bf 68}
(1992)1; B. J. Gluckman, M. L. Span, W.M. Yang, M. Z. Ding, V. In and W. L.
Ditto, Phys. Rev. {\bf E 55} (1997) 4935.

\bibitem{Gills}  Z. Gills, C. Iwata, and R. Roy, Phys. Rev. Lett. {\bf 69}
(1992) 3169.

\bibitem{Bielawski}  S. Bielawski, M. Bouazaoui, D. Derozier, and P.
Glorieux, Phys. Rev. {\bf A47} (1993) 3276.

\bibitem{Zhao1}  H. Zhao, J. Yan, J. Wang, and Y. H. Wang, Phys. Rev. {\bf %
E53} (1996) 299.

\bibitem{Shinbrot}  T. Shinbrot, E. Ott, C. Grebogi, and J. A. Yorke, Phys.
Rev. Lett. {\bf 65} (1990) 3215; E. Bollt and J.D. Meiss, Physica {\bf D81},
(1995) 280; D.J. Christini and J. J. Collins, {\it ibid.} {\bf 53} (1996)
R49.

\bibitem{Zhao2}  H. Zhao, Y. H. Wang, and Z. B. Zhang, Phys. Rev. {\bf E57}
(1998) 5358.

\bibitem{Segard}  B. S\'{e}gard, B. Macke, L.A. Lugiato, F. Prati, and M.
Brambilla, Phys. Rev. {\bf A39} (1989) 703.

\bibitem{Gibbs}  H. M. Gibbs, F. A. Hopf, D. L. Kaplan, and R. L. Shoemaker,
Phys. Rev. Lett. {\bf 46} (1981) 474. F. A. Hopf, D. L. Kaplan, H. M. Gibbs,
and R. L. Shoemaker, Phys. Rev. {\bf A25} (1982) 2172.
\end{references}
\end{document}